\documentclass{epl}

\title{Typical medium theory of Anderson localization: A local order parameter approach to strong disorder effects}
\author{V. Dobrosavljevi\' c\inst{1} \and A.A. Pastor\inst{1} \and  Branislav K. Nikoli\' c\inst{2}}
\institute{
  \inst{1} Department of Physics and National High Magnetic Field Laboratory, 
Florida State University, Tallahassee, FL 32306, U.S.A. \\
  \inst{2} Department of Physics, Georgetown University, Washington, DC 20057-0995, U.S.A.
}

\begin{document}

\maketitle

\begin{abstract}
We present a self-consistent theory of Anderson localization that yields a simple  
algorithm to obtain  \emph{typical local density of states} as an 
order parameter, thereby reproducing  the essential features of a phase-diagram of 
localization-delocalization quantum phase transition in the standard lattice models 
of disordered electron problem. Due to the local character of our theory, it can easily 
be combined with dynamical mean-field approaches to strongly correlated electrons, thus 
opening an attractive avenue for a genuine {\em non-perturbative} treatment of the 
interplay of strong interactions and strong disorder. 
\end{abstract}

After more than four decades of vigorous efforts, there remains little doubt
that disorder-driven metal-insulator transitions (MITs) bear many
similarities to more familiar critical phenomena. The basic physical process
involved was identified by Anderson~\cite{anderson}, who first discussed the
localization of electronic wave functions as a driving force behind such
MITs. Further theoretical progress has been slow, partly due to
ambiguities in identifying an appropriate order parameter for Anderson
localization. Nevertheless, important information was obtained by using
scaling approaches based on $2+\epsilon$ expansions~\cite{gang4}, which were
subsequently extended~\cite{fink} to incorporate the interaction effects.

There are several reasons why the existing theories remain unsatisfactory.
Most importantly, the MITs generically take place at \emph{strong} disorder
where the energy scales associated with both disorder and the interactions
are comparable to the Fermi energy, in contrast to what happens in
perturbative $2+\epsilon$-expansion approaches. As a result, well defined
\emph{precursors} of the MITs are seen even at very high temperatures, as
experimentally demonstrated in many systems~\cite{lr,kravchenko}. These
features include not only the scaling behavior of various quantities, but
also the breakdown of the Matthiessen rule and the Mooij correlation~\cite{lr}. 
To understand such global behavior, a mean-field like formulation would be 
advantageous, but it should be one that can incorporate both the strong disorder (i.e., 
Anderson localization) and the strong correlation effects on the same footing.
\begin{figure}
\onefigure[width=9cm,height=7.5cm]{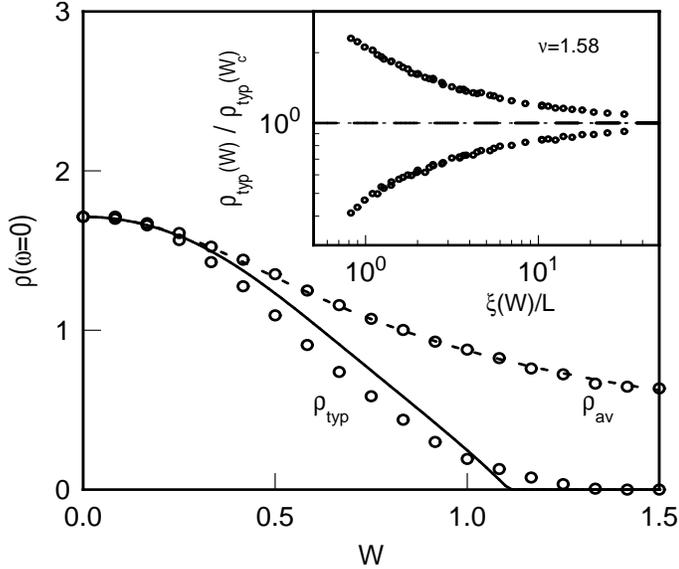}
\caption{Typical and average DOS as a function of disorder $W$, for a 
three-dimensional cubic lattice at the band center ($\protect\omega =0$). Results
from exact numerical calculations (circles) are compared to the predictions
of TMT (for TDOS - full line ) and CPA (for ADOS - dashed line). Finite size
scaling of the numerical data in the critical region $W=1.17-1.58$, and
sizes $L=4-12$ is shown in the inset, where $\protect\rho_{\mathrm{typ}%
}(W,L)/\protect\rho_{\mathrm{typ}}(W_c ,L)$ is plotted as a function of $%
\protect\xi (W)/L$, and $\protect\xi (W) = 0.5 |(W_c -W)/W_c|^{-\protect\nu}$
is the correlation length in units of the lattice spacing. The numerical
data are consistent with $\protect\beta =\protect\nu=1.58$.}
\label{fig:figure3}
\end{figure}

It has been known for a long time that naive mean-field approximation to effective 
field theories of localization (nonlinear $\sigma$-models) leads to apparent order 
parameter that is related to to the density of states, and therefore does not reflect 
critical behavior~\cite{mirlin} (conceptually, this is due to the fact that localization 
manifests in transport quantities, such as the two-particle Green function, rather than 
in the one-particle Green function). This impedes, usually straightforward, implementation 
of the standard concepts of critical phenomena theory, such as order parameter and upper 
critical dimension $d_c$. Nevertheless, the ability of the order parameter to label different 
phases, thereby delineating a complete phase diagram, makes it the most appealing and 
robust outcome of the mean-field philosophy, which survives its quantitative inaccuracy in 
dimensions  below $d_c$ (there is a plethora of evidences that $d_c \rightarrow \infty$ for Anderson 
localization~\cite{mirlin}). In this letter, we demonstrate how an appropriate local order parameter 
can be defined and self-consistently calculated, producing a ``mean-field'' (in the {\em unconventional} 
sense elaborated above) description of Anderson localization. This formulation is \emph{not} restricted 
to either low temperatures or to Fermi liquid regimes. The promising feature of our ``solution'', 
which we term \emph{typical medium theory} (TMT), to the order parameter puzzle, is in its 
potential usefulness outside of the realm of {\em pure} Anderson localization of {\em non-interacting} 
particles. Being amenable to easy incorporation  into the well-known dynamical mean-field theories 
(DMFT)~\cite{dmfrmp,dmftdis} of strongly correlated systems, TMT offers a novel framework to address 
questions  difficult to tackle by any alternative formulation, but which are of crucial importance 
for many electron systems of current interest dominated by non-perturbative physics. 

Our starting point is motivated by the original formulation of Anderson~\cite{anderson}, 
which adopts a \emph{local} point of view~\cite{local}, and
investigates the possibility for an electron to \emph{delocalize} from a
given site at large disorder. This is most easily accomplished by
concentrating on the (unaveraged) local density of electronic states (LDOS) 
$\rho_i (\omega ) =\sum_n \delta (\omega -\omega_n ) |\psi_n (i)|^2$. In contrast 
to the global (averaged) density of states (ADOS) which is not critical at the Anderson 
transition, LDOS undergoes a qualitative change upon localization, as first noted in 
Ref.~\cite{anderson}. This follows from the fact that LDOS directly measures the local 
amplitude of the electronic wave functions. When electrons localize, the local spectrum turns from a
continuous to an essentially discrete one~\cite{anderson}, but the \emph{%
typical} value of LDOS vanishes. Just on the metallic side, but very close
to the transition, these delta-function peaks turn into long-lived resonance
states and thus acquire a finite \emph{escape rate} from a given site.
According to Fermi's golden rule, this escape rate can be estimated~\cite{anderson} 
as $ \tau_{\mathrm{esc}}^{-1}\sim t^2\rho$, where $t$ is the
inter-site hopping element, and $\rho$ is the density of local states of the
immediate neighborhood of a given site.

The \emph{typical} escape rate is thus determined by the typical local
density of states (TDOS), so that TDOS directly determines the conductivity
of the electrons. This simple argument strongly suggests that TDOS should be
recognized as an appropriate order parameter for Anderson localization.
Because the relevant distribution function for LDOS becomes increasingly
broad as the transition is approached, the desired typical value is well
represented by the \emph{geometric average} $\rho_{\mathrm{typ}} = \exp\{
\langle\ln \rho \rangle\} $, where $\langle \cdots\rangle$ represents the
average over disorder. Interestingly, recent scaling analysis~\cite{janssen}
of the multifractal behavior of electronic wave functions near the Anderson
transition have independently arrived at the same conclusion, identifying
TDOS as defined by the geometric average as the viable candidate for an order 
parameter (somewhat related ideas have also been discussed earlier in 
Ref.~\cite{logan}). A complementary insight comes from effective 
field theories, applied to special models (e.g., localization on the Bethe lattice) that 
can be solved in the strong coupling limit, which point out that the full LDOS 
distribution should be considered an an ``order parameter function''~\cite{mirlin}. 
Despite these advances, a transparent technique to obtain order parameter within the 
framework of standard models of localization (such as tight-binding Hamiltonian with 
diagonal disorder---the so-called Anderson model that is usually a non-interacting 
disordered piece of lattice Hamiltonians of strongly correlated fermions) is lacking. 
Our principal result is shown in Fig.~\ref{fig:figure3}, where comparison of the phase 
diagram obtained from TMT to the numerically exact treatment of the Anderson model on a 
3D simple cubic lattice demonstrates that such approach is fully capable of capturing 
essential features (e.g., qualitative positions of the phase boundaries) of 
localization-delocalization transition occurring at strong disorder in dimensions  
$d > 2$. Thus, despite being an uncontrollable approximation that does not reproduce correct 
critical exponents in any dimensionality, TMT paves the way to include the effects of strong 
disorder into modern DMFT approaches to strongly correlated systems, where it has been a 
daunting task to go beyond coherent-potential approximation (CPA)~\cite{cpa} (CPA cannot 
account for quantum interference effects induced by scattering off impurities, i.e., 
localization).

To formulate a self-consistent theory for our order parameter, we follow the
``cavity method,'' a general strategy that we borrow from the DMFT~\cite{dmfrmp}. 
In this approach, a given site is viewed as being embedded in an
effective medium characterized by a local self energy function $\Sigma
(\omega )$. For simplicity, we concentrate on a single band tight-binding
model of noninteracting electrons with random site energies $\varepsilon_i$
with a given distribution $P( \varepsilon_i)$. The corresponding local
Green function then takes the form $G (\omega , \varepsilon_i) = [\omega -\varepsilon_i -\Delta (\omega )]^{-1}$, 
where the ``cavity function'' is given by $\Delta(\omega )= \Delta_o (\omega -\Sigma(\omega ))$ with 
$\Delta_o (\omega ) = \omega -1/G_o (\omega )$. The lattice Green function $G_o (\omega ) = 
\int_{-\infty}^{+\infty}d\omega ^{\prime}\; \frac{D (\omega ^{\prime})}{\omega -\omega ^{\prime}}$ 
is the Hilbert transform of the bare density of states $D (\omega )$. Given the effective medium specified 
by a self-energy $\Sigma (\omega )$, we are now in the position to evaluate the order parameter, which we choose to
be TDOS as given by
\begin{equation}\label{eq:six}
\rho_{\mathrm{typ}} (\omega )= \exp\left\{ \int d\varepsilon_i \;
P(\varepsilon_i)\; \ln \rho (\omega , \varepsilon_i)\right\},
\end{equation}
where LDOS $\rho (\omega , \varepsilon_i)= - \mathrm{Im} \, G (\omega,\varepsilon_i)/\pi$. To 
obey causality, the Green function corresponding to $\rho_{\mathrm{typ}} (\omega )$ must be
specified by analytical continuation, which is performed by the Hilbert
transform
\begin{equation}\label{eq:seven}
G_{\mathrm{typ}} (\omega ) = \int_{-\infty}^{+\infty}d\omega ^{\prime}\;
\frac{\rho_{\mathrm{typ}} (\omega ^{\prime})}{\omega -\omega ^{\prime}}.
\end{equation}
Finally, we close the self-consistency loop by setting the Green functions
of the effective medium to be equal to that corresponding to the local order
parameter, so that $G_{\mathrm{em}} (\omega )=G_o (\omega -\Sigma (\omega ))=G_{\mathrm{typ}}
(\omega )$. It is important to emphasize that the procedure defined by these equations  
is not specific to the problem at hand. The same strategy can be used in any theory
characterized by a {\em local self-energy}. The only requirement specific to our
problem is the definition of TDOS as a local order parameter given by Eq.~(\ref{eq:six}). 
If we choose the \emph{algebraic} instead of the {\em geometric} average of
LDOS, our theory would reduce to CPA~\cite{cpa}, which produces excellent results for the
ADOS for any value of disorder, but finds no Anderson transition. Thus TMT
is a theory having a character very similar to CPA, with a small but crucial difference---the 
choice of the correct order parameter for Anderson localization.
\begin{figure}
\onefigure[scale=0.45]{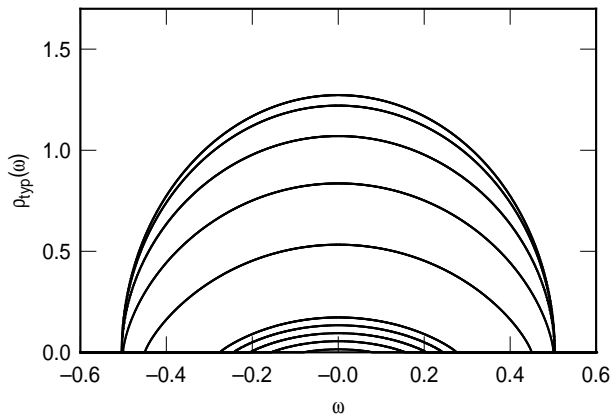}
\caption{Typical density of states for the SC model, for disorder values $W$
= 0, 0.25, 0.5, 0.75, 1, 1.25, 1.275, 1.3, 1.325, 1.35. The entire band
localizes for $W=W_c =e/2\approx 1.359$.}
\label{fig:figure1}
\end{figure}

\begin{figure}
\onefigure[scale=0.44]{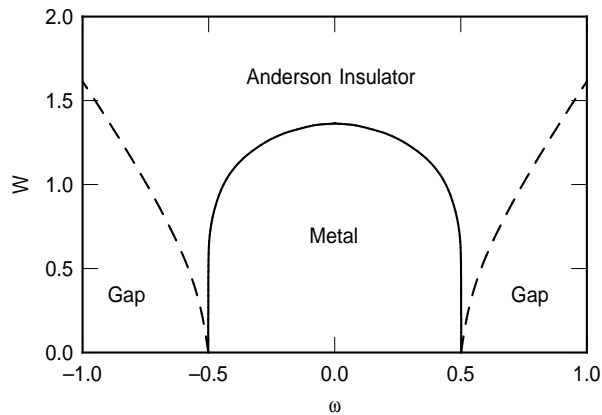}
\caption{Phase diagram for the SC model. The trajectories of the mobility
edge (full line) and the CPA band edge (dashed line) are shown as a function
the disorder strength $W$.}
\label{fig:figure2}
\end{figure}

In our formulation, as in DMFT, all the information about the electronic
band structure is contained in the choice of the bare DOS $D (\omega )$. It
is not difficult to solve the above equations numerically, which can be efficiently
done using FFT methods~\cite{dmfrmp}. We have done so for several model
densities of states, and find that most of our qualitative conclusions do
not depend on the specific choice of band structure. We illustrate these
findings using a simple semicircular (SC) model for the bare DOS given by $D
(\omega ) = \frac{4}{\pi}\sqrt{1-(2\omega )^2}$, for which $\Delta_o (\omega
) =G_o (\omega )/16$~\cite{dmfrmp}. Here and in the rest of the paper all
the energies are expressed in units of the bandwidth, and the random site
energies $\varepsilon_i$ are uniformly distributed over the interval $[-W/2
, W/2]$. The evolution of TDOS as a function of $W$ is shown in Fig.~\ref{fig:figure1}. The
TDOS is found to decrease and eventually vanish even at the band center for $%
W=W_c\approx 1.36$. When $W < W_c$, the part of the spectrum where TDOS
remains finite corresponds to the region of extended states, and is found to
shrink with disorder, indicating that the band tails begin to localize. The
resulting phase diagram is presented in Fig.~\ref{fig:figure2}, showing the trajectories of
the mobility edge (as given by the frequency where TDOS vanishes for a given
$W$), and the band edge (where the ADOS calculated by CPA vanishes).

Further insight in the critical behavior is obtained by noting that near $%
W=W_c$ it proves possible to analytically solve TMT equations. 
Concentrating for simplicity on the band center ($\omega =0$), we can expand these equations  
in powers of the order parameter $\rho_o =\rho_{\mathrm{typ}} (0)$ giving $\rho_o = a\rho_o - b\rho_o^2 +\cdots$, where 
 $a=\exp\left\{ -2\int d\varepsilon P(\varepsilon )\ln |\varepsilon |\right\}$ and $b=2aP(0)$. 
The transition where $\rho_o$ vanishes is found at $a=1$, giving $W=W_c =e/2=1.3591$, consistent with our 
numerical solution. Near the transition, to leading order $\rho_o (W) = \left(\frac{4}{\pi}\right)^2 (W_c -W)$, 
meaning that the order parameter exponent is $\beta =1$. The analytical solution is more difficult to 
obtain for arbitrary $W$. Still, the above approach can be extended to find a full frequency-dependent
solution close to the critical value of disorder $W=W_c$. There it assumes a
simple scaling form $\rho_{\mathrm{typ}}(\omega ,W) = \rho_o (W) f\left[\omega /\omega_o
(W)\right]$,  with $\omega_o (W)=\sqrt{\left(\frac{e}{4}\right)(W_c -W)}$ and the 
scaling function $f(x)= 1-x^2$. This is again consistent in detail with the numerical
solution of Fig.~\ref{fig:figure1} (note that TDOS curves assume a simple \emph{parabolic}
shape close to $W=W_c$).

In order to examine the quantitative accuracy of out theory, we have carried
out exact numerical calculations for a three-dimensional cubic lattice with
random site energies, using Green functions for an open finite sample
attached to two semi-infinite clean leads~\cite{nikolic}. We have computed
both the average and the typical DOS at the band center as a function of
disorder, for cubes of sizes $L$ =4, 5, 6, 7, 8, 9, 10, 11, and 12, and
averages over 1000 sample realizations, in order to obtain reliable data by
standard finite size scaling procedures. The TMT and CPA equations for the
same model were also solved by using the appropriate bare DOS (as expressed
in terms of elliptic integrals), and the results are presented in 
Fig.~\ref{fig:figure3}. We find remarkable agreement between the numerical 
data and the self-consistent CPA calculations for the ADOS, but also a surprisingly good
agreement between the numerical data and the TMT predictions for the TDOS
order parameter. For a cubic lattice, the exact value is $W_c \approx 16.5/12=1.375$~\cite{slevin}, 
whereas TMT predicts a 20\% smaller value $W_c\approx 1.1$. The most significant 
discrepancies are found in the critical region, since TMT predicts the order 
parameter exponent $\beta=1$, whereas the exact value is believed to be 
$\beta\approx 1.58$~\cite{exponents}, consistent with our numerical data. 
Nevertheless, we conclude that TMT is as accurate as one can expect from a 
simple ``mean-field'' like formulation~\cite{symmetries}.

Next, we address the transport properties within TMT. The escape rate from a
given site can be rigorously defined in terms of the cavity field $\Delta(\omega)$, 
and using our solution of the TMT equations, we find $\tau_{\mathrm{esc%
}}^{-1} = -\mathrm{Im}\, \Delta (0) \sim \rho_{\mathrm{typ}} \sim (W_c -W )$.
To calculate the conductivity within our local approach, we follow a
strategy introduced by Girvin and Jonson~\cite{girvin}, who pointed out that
close to the localization transition the conductivity can be expressed as $%
\sigma = \Lambda \langle A_{12}A_{21}-A_{11}A_{22}\rangle$, where $A_{ij}=- 
\mathrm{Im}\, G_{ij}$ is the spectral function corresponding to the nearest 
neighbor two-site cluster. We have computed $a_{12}$ by examining two sites 
embedded in the effective medium defined by TMT, thus allowing for localization 
effects. The vertex function $\Lambda$ remains~\emph{finite} at the localization 
transition~\cite{girvin}, and thus can be computed within CPA~\cite{lambda}. 
The resulting critical behavior of the $T=0$ conductivity follows that of 
the order parameter, $\sigma \sim \rho_{\mathrm{typ}}\sim (W_c -W)$, giving 
the conductivity exponent $\mu$ equal to the order parameter exponent $\beta$, 
consistent with what is expected~\cite{exponents}.
\begin{figure}
\onefigure[scale=0.45]{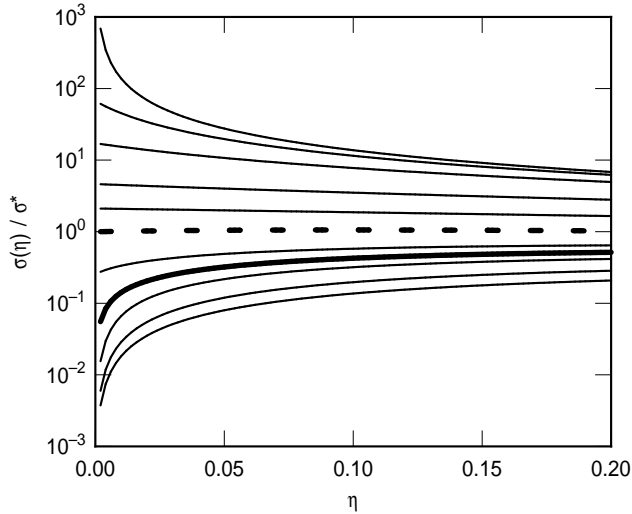}
\caption{Conductivity as a function of the inelastic scattering rate $%
\protect\eta$ for the SC model at the band center and $W$ = 0, 0.125, 0.25,
0.5, 0.75, 1, 1.25, 1.36, 1.5, 1.75, 2. The ``separatrix'' ($\protect\sigma %
= \protect\sigma^*$ independent of $\protect\eta$, i. e. temperature) is
found at $W=W^*\approx 1$ (dashed line). The critical conductivity $\protect%
\sigma_c (\protect\eta )\sim \protect\eta^{1/2}$ corresponds to $W=W_c =1.36$
(heavy full line).}
\label{fig:figure4}
\end{figure}
Finally, we examine the temperature dependence the conductivity as a
function of $W$. Physically, the most important effect of finite
temperatures is to introduce finite inelastic scattering due to interaction
effects. At weak disorder, such inelastic scattering increases the
resistance at higher temperatures, but in the localized phase it produces
the opposite effect, since it suppresses interference processes and
localization. To mimic these inelastic effects within our noninteracting
calculation, we introduce by hand an additional scattering term in our
self-energy, viz. $\Sigma \rightarrow \Sigma -i\eta$. The parameter $\eta$
measures the inelastic scattering rate, and is generally expected to be a
monotonically increasing function of temperature. The resulting behavior of
the conductivity as a function of $\eta$ and $W$ is presented in Fig.~\ref{fig:figure4}. 
As  $\eta$ (i.e., temperature) is reduced, we find that the conductivity curves
``fan out'', as seen in many experiment close to the MIT~\cite{lr,kravchenko}. 
Note the emergence of a ``separatrix''~\cite{lr,kravchenko} where the
conductivity is temperature independent, which is found for $W\approx 1$,
corresponding to $k_F \ell \sim 2$, consistent with some experiments~\cite{lr}. 
At the MIT, we find $\sigma_c (\eta )\sim \rho_{\mathrm{typ}} (\eta )\sim 
\eta^{1/2}$.

In summary, we have formulated a local order parameter theory for
disorder-driven MITs that in absence of interactions reproduces most of the
expected features of the Anderson transition. In addition, the role of
strong electronic correlations near disorder-driven MITs can be readily
examined within the typical medium theory, since the local character of our
approach offers a natural starting point for incorporating both the
localization and the interaction effects using the DMFT framework.

\acknowledgments
We thank S. Chakravarty, S. Das Sarma, J. K. Freericks, D. A. Huse, G.
Kotliar, E. Miranda, D. Popovi\'c, S. H. Simon, and S. Sondhi for useful
discussions. This work was supported by the NSF grant DMR-9974311 and the
National High Magnetic Field Laboratory (VD and AAP), and the ONR grant
N00014-99-1-0328 (BKN).


\begin{thebibliography}{0}
\bibitem{local} Ours is a diametrically opposite, but complementary approach
to that of the weak localization theories~\cite{lr}, which investigate the
stability of the metallic phase with respect to weak impurity scattering.
The self-consistent theory of localization of Vollhardt and W\"olfle~\cite{vw} 
has a similar character, since it essentially represents a resummation
of the most singular contributions at weak disorder near two dimensions. On
the other hand, some recent experiments~\cite{kravchenko} indicate~\cite{gang2} 
that the observed quantum critical behavior may be dominated by the physics of 
the insulator, thus favoring the local picture.
\bibitem{exponents} Arguments based on the multifractal scaling analysis~\cite{janssen}, 
indicate that the conductivity exponent $\mu =(d-2)\nu$, and the order parameter 
exponent $\beta = (\alpha_o -d)\nu$, where $\nu$ is the correlation length exponent, 
and $\alpha_o$ characterizes the multifractal spectra of wave function amplitudes. In 
addition, numerical calculations find that in three dimensions, $\alpha_o \approx 
4$~\cite{schreiber}, and $\nu \approx 1.58$~\cite{slevin}, so that 
$\beta \approx \mu \approx 1.58$. The result $\beta =\mu = 1 + O(\epsilon)$ 
is also found within the $2+\epsilon$ approach~\cite{janssen}.
\bibitem{symmetries} Within a ``mean-field'' like formulation such as ours, we
do not expect the different symmetry classes (with respect to time-reversal
or spin-rotation invariance) to result in different critical exponents.
Although it is well known that these differences play a crucial role in the
weak-localization regime~\cite{lr}, we note that recent numerical 
work~\cite{symplectic} finds very similar critical behaviors in 
three dimensions in all cases.
\bibitem{lambda} We have used the CPA approach to evaluate the vertex
function as $\Lambda = \sigma_{cpa}/a_{12}^{cpa}$, where $\sigma_{cpa}$ is
the CPA conductivity~\cite{cpa}, and $a_{12}^{cpa}$ is the two-site
correlation function embedded in the CPA effective medium. Since TMT 
coincides with CPA for weak disorder, our results reduce to the correct
value at $W << W_c$, where the conductivity assumes the Drude-Boltzmann form.
\end{thebibliography}
\end{document}